\documentstyle[gcdv,psfig]{article}

\def\reference{\parskip 0pt\par\noindent\hangindent 0.5 truecm}

\newcommand{\lsim}{\raisebox{-3.8pt}{$\;\stackrel{\textstyle <}{\sim}\;$}}

\begin{document}

\small
\shorttitle{M/L ratio, IMF and chemical evolution in disc galaxies}
\shortauthor{L.\ Portinari et~al.}

\title
{\large \bf
Mass--to--Light ratio, Initial Mass Function and chemical evolution 
in disc galaxies
}
%

\author{\small 
L.~Portinari$^{1}$, J.~Sommer--Larsen$^{1}$, R.~Tantalo$^{2}$
} 

\date{}
\twocolumn[
\maketitle
\vspace{-20pt}
\small
{\center
$^1$ Theoretical Astrophysics Center, Juliane Maries Vej 30, DK-2100
Copenhagen \O, Denmark\\ 
$^{2}$ Dipartimento di Astronomia, Universit\`a di Padova, Vicolo 
dell'Osservatorio 2, I-35122 Padova, Italy \\
e--mail: lportina, jslarsen@tac.dk; tantalo@pd.astro.it \\[3mm]
}

\begin{center}
{\bfseries Abstract}
\end{center}
\begin{quotation}
\begin{small}
\vspace{-5pt}
\noindent
Cosmological simulations of disc galaxy formation, when compared
to the observed Tully--Fisher relation, suggest a low Mass--to--Light (M/L)
ratio for the stellar component in spirals. We show that a number of
``bottom--light'' Initial Mass Functions (IMFs) suggested independently in
literature, do imply M/L ratios as low as required, at least for late
type spirals (Sbc/Sc). However the typical M/L ratio, and correspondingly 
the zero--point of the Tully--Fisher relation, is expected to vary considerably
with Hubble type. 

Bottom--light IMFs tend to have a metal production in excess of what
is tipically estimated for spiral galaxies. Suitable tuning of the IMF slope
and mass 
limits, post--supernova fallback of metals onto black holes or metal outflows
must then be invoked, to reproduce the observed chemical properties of disc
galaxies.

\end{small}
\end{quotation}
]

\begin{figure}
{\centerline{ \psfig{file=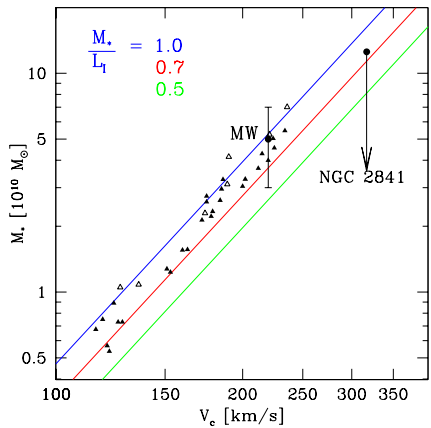,width=5.9truecm} }}
\caption{Straight lines: observed Tully--Fisher relation for 
Sbc--Sc disc galaxies (Dale et~al.\ 1999; h=0.7 adopted here), 
for different assumptions about the stellar M/L$_I$ ratio. Triangles:
simulated galaxies by Sommer--Larsen et~al.\ (2003).
Also shown is the location of the Milky Way and NGC 2841.}
\label{fig:tully}
\end{figure}

\section{Introduction}

Recent N--body+SPH cosmological simulations of the formation of disc galaxies 
reproduce the observed Tully-Fisher (TF) relation (Dale et~al.\ 1999),
provided the mass--to--light (M/L) ratio of the stellar component is rather 
low, {\mbox{M/L$_I$=0.7--1}} in the I--band 
(Sommer-Larsen \& Dolgov 2001; Sommer-Larsen et~al.\ 2003; 
Fig.~\ref{fig:tully}). The location of the simulated galaxies in the
($M_*$, $V_c$) plane of Fig.~\ref{fig:tully} is quite independent of 
the adopted Initial Mass Function (IMF) or feedback efficiency: the baryonic 
mass that cools out 
to form a galactic disc and its resulting circular velocity correlate
so that data points tend to move along the TF relation,
hardly affecting the zero--point (Navarro \& Steinmetz 2000ab). 
However, the IMF is crucial for the M/L ratio, to translate the stellar masses
$M_*$ to luminosities and compare the simulated TF relation
to the empirical one. Although the zero--point of the simulated TF
may change with the concentration of the dark matter halos,
and hence with the normalization of the power spectrum $\sigma_8$
(Navarro \& Steinmetz 2000ab; Eke et~al. 2001), many other arguments 
support a low stellar M/L ratio in spiral galaxies:

The stellar mass of the Milky Way is 
{\mbox{M$_* \sim 5 \times 10^{10}~M_{\odot}$}}; to lie on the observed TF 
relation as other spirals, its M/L$_I$ must be $\lsim$1 (Sommer-Larsen 
\& Dolgov 2001; Fig.~\ref{fig:tully}).
A low {\mbox{M/L$_I < 0.8$}} is also derived for the massive
Sb galaxy NGC 2841, 
when compared to the observed TF relation (Portinari et~al.\ 2004a, hereinafter
PST; Fig.~\ref{fig:tully}).

Based on bar instability arguments, Efstathiou et~al.\ (1982) suggest an 
upper limit of {\mbox{M/L$_B \leq 1.5~h$}} for discs, i.e.\ M/L$_B \lsim 1$ 
for $h$=0.7.
($h$ indicates the Hubble constant H$_0$ in units of
100~km~sec$^{-1}$~Mpc$^{-1}$).

The stellar M/L ratio is related to the issue as to whether discs are maximal
or sub--maximal, i.e.\ as to whether they dominate or not the dynamics 
and rotation curves in the inner galactic regions.
Even in the case of maximal stellar discs, lower M/L ratios
for the stellar component are required, than those predicted by
the Salpeter IMF (Bell \& de Jong 2001). And it is still much debated
whether discs are maximal or sub--maximal;
for his favoured sub--maximal disc model, Bottema (2002) finds
M/L$_I \sim 0.82$.

Finally, two recent dynamical studies of individual spiral galaxies yield 
{\mbox{M/L$\sim$1}}
in the B, V and I band for the Sc galaxy NGC 4414 (Vallejo et~al.\ 2002)
and M/L$_I$=1.1 for the disc of the Sab spiral 2237+0305, 
Huchra's lens (Trott \& Webster 2002).

In this paper we discuss if M/L ratios so low are compatible with our 
understanding of stellar populations and chemical evolution in 
disc galaxies. We also address the effects of different star formation 
histories
on the TF relation for different Hubble types.

\section{Star Formation History \\ 
and Initial Mass Function}

\noindent
The M/L ratio of the stellar component of a galaxy (including both living stars
and remnants) depends on the stellar Initial Mass Function (IMF) and on
the star formation history (SFH) of the system. 

\begin{figure}
{\centerline{ \psfig{file=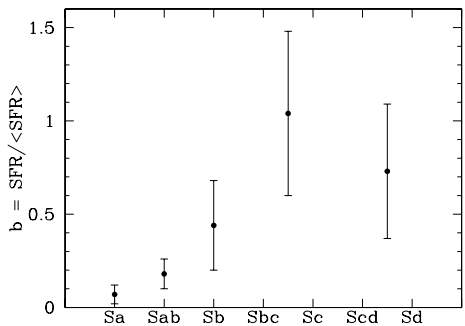,angle=270,width=6.4truecm} }}
\caption{Mean $b$--parameter as a function of Hubble type. Data from 
Sommer--Larsen et~al.\ (2003).}
\label{fig:btype}
\end{figure}

The SFH of a disc galaxy is related to its Hubble type:
Kennicutt et~al.\ (1994) demonstrated that the sequence of spiral types 
is in fact a sequence of different SFHs in the discs, as traced by the 
birthrate parameter
\[ b = \frac{SFR}{<SFR>} \]
or the ratio between the present and the past average star formation rate 
(SFR); see Fig.~\ref{fig:btype}.
Our reference TF relation by Dale et~al.\ (1999) in Fig.~\ref{fig:tully} is 
representative for Sbc--Sc spirals (Giovanelli et~al.\ 1997).
Kennicutt et~al.\ find that Sbc--Sc discs correspond to $b$=0.8--1, hence
we will concentrate our discussion on objects with SFHs corresponding to
these values for $b$.

As to the IMF, a standard Salpeter slope extended over the mass range 
[0.1--100]~$M_{\odot}$ certainly yields much higher M/L ratios than those 
mentioned in the introduction (Fig.~\ref{fig:b-models}, top left). 
There is however plenty of observational evidence that the IMF
flattens below $\sim 1 M_{\odot}$, possibly 
with a turn--over at low masses, and hence is ``bottom--light'' with respect
to the Salpeter IMF.

\begin{figure}
{\centerline{ \psfig{file=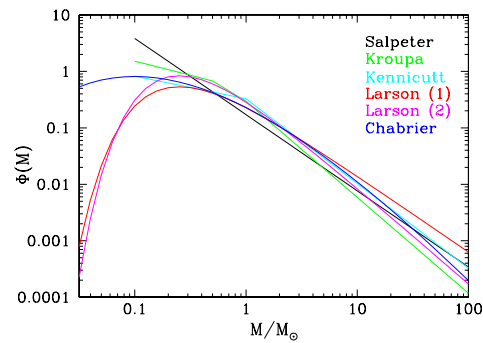,angle=270,width=6.4truecm} }}
\caption{Comparison between the Salpeter IMF and the ``bottom--light'' IMFs 
considered in this paper, all normalized to the same total integrated mass 
of 1~$M_{\odot}$.}
\label{fig:IMF}
\end{figure}

We considered the following IMFs:
the {\bf Salpeter} (1955) IMF;
the {\bf Kroupa} (1998) IMF, derived from field stars in the Solar 
Neighbourhood;
the {\bf Kennicutt} IMF, derived from the global 
properties of spiral galaxies (Kennicutt et~al.\ 1994);
the {\bf Larson} (1998) IMF, with an exponential cut--off 
at low masses as favoured by recent determinations of the local IMF 
down to the sub--dwarf regime (Chabrier 2001, 2002);
a {\bf modified Larson} IMF, with the same peak mass but a steeper slope 
at high masses, in accordance with local determinations (Scalo 1986);
the {\bf Chabrier} (2001, 2002) IMF, derived from low mass stars and brown 
dwarfs in the local field. 
With respect to Salpeter, all the other IMFs are ``bottom--light'' 
(Fig.~\ref{fig:IMF}). 

\section{Results from simple models}

We will focus on the M/L ratio in the I--band, 
which is an excellent optical tracer of stellar mass, because
the I--band luminosity is less sensitive to recent sporadic star formation 
and to corrections for dust extinction than bluer bands. Besides,
the M/L ratio of Single Stellar Populations (SSPs) in the I--band 
is less metallicity dependent 
than in other bands, and it is less sensitive to the specific treatment 
of the AGB phase than redder, NIR bands (PST).

\begin{figure*}
{\centerline{ \psfig{file=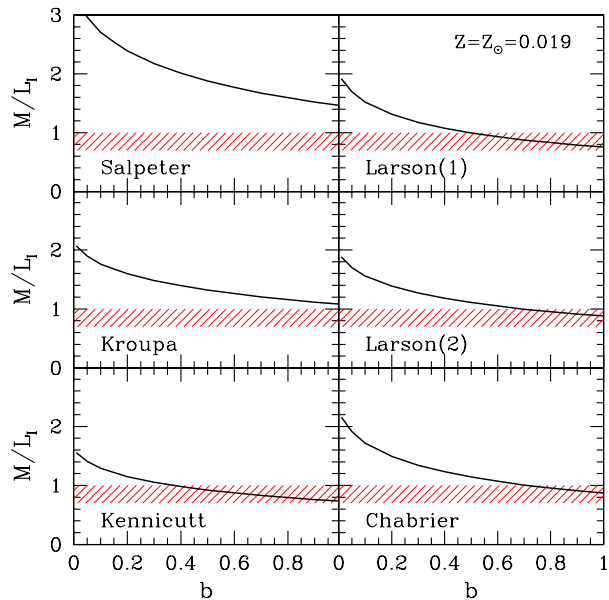,width=8.4truecm} }}
\caption{I--band M/L ratio at varying $b$--parameter of the SFH, for different
IMFs. The red shaded area marks the range
M/L$_I$=0.7--1 favoured by observations for Sbc--Sc discs (corresponding
to $b$=0.8--1).}
\label{fig:b-models}
\end{figure*}

We computed SSPs based on the latest Padua isochrones 
(Girardi et~al.\ 2002), for the different
IMFs in \S 2. These are to be convolved with suitable SFHs.
We generated a set of SFHs characterized by different values of the 
$b$--parameter, by adopting an exponentially decaying 
{\mbox{$SFR \propto e^{- \frac{t}{\tau}}$}} with different decaying rates.
SSP metallicities around solar can be considered typical for spiral 
galaxies. 
The resulting M/L ratio of the global stellar population
(including remnants) as a function of the {\mbox{$b$--parameter}}
is displayed in Fig.~\ref{fig:b-models}, for the different IMFs. 
The range $0.8 \leq b \leq 1$ is representative of Sbc--Sc spirals.
The red shaded area indicates the range in M/L=0.7--1 suggested in \S 1.

Fig.~\ref{fig:b-models} shows that bottom--light IMFs can in fact yield 
M/L$_I < 1$ for 
late--type spirals, though one probably needs slightly ``lighter'' IMFs than 
the local Kroupa one.

\section{Offsets of the TF relation with Hubble type}

From Fig.~\ref{fig:b-models}, the stellar M/L ratio is expected to vary 
with Hubble type due to the differences in SFH parameterized by $b$. 
This effect implies systematic offsets with Hubble 
type of the luminosity zero--point of the TF relation (Rubin et~al.\ 1985; 
Giovanelli et~al.\ 1997; Kannappan et~al.\ 2002). 

\begin{figure}
{\centerline{ \psfig{file=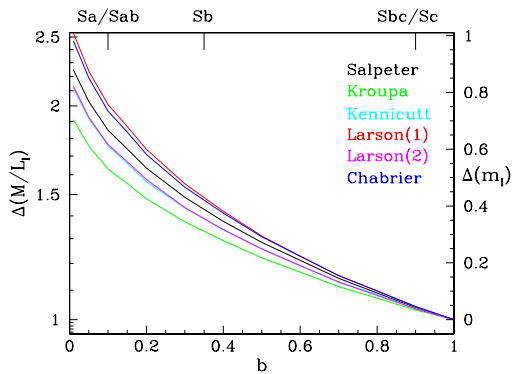,angle=270,width=6.7truecm} }}
\caption{Relative M/L ratio normalized to the value corresponding to $b$=1 
models.}
\label{fig:TFoffsets}
\end{figure}

Fig.~\ref{fig:TFoffsets} shows the M/L ratio as a function of $b$, normalized
to the value corresponding to $b$=1. The scale on the right axis indicates 
the corresponding shift in magnitude. With respect to Sbc/Sc spirals,
we predict a systematic offset of 0.3--0.4~mag for Sb's 
($b \sim$0.35) and of 0.6--0.8~mag for Sa/Sab's ($b \sim$0.1),
as a result of the different characteristic SFHs.
These offsets are only slightly reduced when bulges are added to discs in the 
computation of the total M/L ratios of galaxies (PST).
The offsets we predict are larger than the empirical ones found by Giovanelli 
et~al.\ (1997): 0.1~mag for Sb spirals and 0.32~mag for earlier types.
However, the extent of the observed offsets in the TF relation is still 
a matter 
of debate: for instance, the larger offsets found by Kannappan et~al.\ (2002) 
in the R--band, 0.76~mag for Sa's, are in good agreement with our predictions.

The characteristic SFH of a galaxy is traced by its colours, so that
the offsets in M/L ratio due to different SFHs can be corrected for, 
by applying suitable M/L vs.\ colour relations (Bell \& de Jong 2001; PST).

\section{Bottom--light IMFs and  chemical evolution}

Besides simple models with exponentially declining SFHs, we also computed
more realistic, multi-zone chemo-photometric models of galactic discs, 
including infall, inside--out formation and radially varying star formation 
efficiency. Chemical evolution 
is followed with the code by Portinari et~al.\ (1998), Portinari \& Chiosi 
(1999) and the models are calibrated to reproduce the typical metallicity 
and metallicity gradient of Sbc/Sc discs (PST). Six sets 
of models have been computed for the six IMFs in \S 2. The corresponding 
photometric properties are calculated by convolving the SFH and metal
enrichment history of each annulus of the disc, with a grid of SSPs of
metallicities between $5 \times 10^{-4} Z_{\odot}$ and $5 Z_{\odot}$ (PST). 

\begin{figure}
{\centerline{ \psfig{file=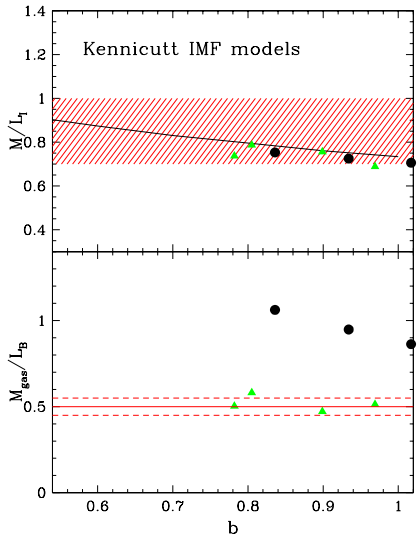,width=5.7truecm} }}
\caption{{\it Top panel}: M/L ratio for chemo--photometric models with the
Kennicutt IMF (solid dots); they confirm the prediction from the simple
exponential models (black line) and are within the observed range
of M/L$_I$=0.7--1 (red shaded area). {\it Bottom panel}:
Predicted gas fractions compared to the observed range, marked by the
red dashed lines. Green triangles: models with IMF mass 
limits tuned between 0.05--0.1~$M_{\odot}$ and 30--35~$M_{\odot}$,
rather than the standard mass range [0.1--100]~$M_{\odot}$, to reproduce
the observed gas fraction.}
\label{fig:models-kenn}
\end{figure}

For each IMF, the chemo--photometric models confirm
the M/L ratios predicted, as a function of $b$, by the simple models 
in Fig.~\ref{fig:b-models}; see the example for the Kennicutt IMF models 
in Fig.~\ref{fig:models-kenn}, top panel.
The results of \S 3 are thus confirmed: the 
``bottom--light'' IMFs considered here imply M/L$_I <$1 for Sbc/Sc spirals,
as required in \S 1. This conclusion remains valid also when the
contribution of the bulge to the global M/L ratio is included (PST).

Chemo--photometric models allow also an insight on the implications
of ``bottom--light'' IMFs for chemical evolution. Some of the IMFs considered
(Kennicutt, Larson and Chabrier) are too efficient in metal production
to reproduce the observed properties of spirals. In particular, the resulting
gas fraction is much larger than the observed
{\mbox{$M_{gas}/L_B \sim 0.5~M_{\odot}/L_{\odot}$}} 
(solid dots in Fig.~\ref{fig:models-kenn}, bottom panel).
For a given IMF and corresponding metal production, the final gas metallicity 
predicted by chemical models increases at decreasing gas fraction (Tinsley
1980, Pagel 1997). Our models, calibrated to reproduce the observed 
metallicities, 
tend to predict
too high gas fractions. Conversely, if they were calibrated to reproduce the
observed gas fractions, they would result in too high metallicities.

This excessive metal production 
is readily understood since the enrichment efficiency of a stellar 
population, or its ``net yield'', is inversely proportional to 
the mass fraction that remains forever locked in low--mass
stars and remnants (Tinsley 1980; Pagel 1997); and for ``bottom--light'' IMFs 
the locked--up fraction is small.
This effect can be compensated by a steep slope at the high-mass end, which
reduces the number of massive stars and the related metal production:
the Kroupa or modified--Larson IMFs, for instance, with a steep
Scalo slope do not overproduce metals (see PST for details). A steep slope 
for the integrated field stars IMF is expected, in fact, if stars form
in star clusters of finite size even from an intrinsically shallower IMF 
(Kroupa \& Weidner 2003).

For those bottom--light IMFs that do imply an excess in metal production,
one way to reconcile models with observations is to tune the upper mass limit 
of the IMF, again reducing the number of massive stars and related metal 
production --- or equivalently assume that above a certain progenitor mass, 
the metals produced by a supernova fall back onto a black hole after the 
explosion. The observed gas fractions can then be matched
without altering the stellar M/L ratio significantly (green triangles 
in Fig.~\ref{fig:models-kenn}).

Alternatively, we need to invoke substantial outflows of metals 
from disc galaxies into the intergalactic medium, to reconcile the high 
enrichment efficiency with 
the observed low gas fractions. This behaviour is reminiscent of that of
elliptical galaxies, responsible for the enrichment of the hot gas in clusters.

With ``standard'' IMFs suited to model the chemical evolution of the Solar 
Neighbourhood (e.g.\ the Kroupa IMF) it is impossible to account for the 
observed metal enrichment in clusters (Portinari et~al.\ 2004b). 
Possibly, some of the ``bottom--light'' IMFs advocated here 
to reproduce low disc M/L ratios, suggest a scenario where the IMF and 
the enrichment efficiency 
may be the same in spiral and cluster galaxies,  and in both cases 
much of the metals are dispersed into the intergalactic medium. 
However, substantial outflows would challenge our understanding of disc 
galaxy formation: in galactic discs, star formation proceeds at a smooth,
non burst--like pace and the observed ``fountains'' and ``chimneys'' 
do not have enough energy to escape the galactic potential; winds are far less
plausible than in spheroids. Moreover, 
strong ongoing stellar feedback and outflows could significantly hamper the 
dynamical formation of galactic discs from the cool--out of halo gas.

The alternative to major winds from disc galaxies is that the metal production
in spirals is much lower than in galaxy clusters, because of 
a different IMF (Portinari et~al.\ 2004b). The IMF may 
change out of Jeans--mass variations with redshift (Moretti et~al.\ 2003 and 
references therein); or be a universal function within star clusters, but
lead statistically to more high--mass stars in massive ellipticals,
where in regimes of intense star formation larger star clusters can be
formed (Kroupa \& Weidner 2003).

\section*{Acknowledgments}
LP is grateful to the organizers and to Swinburne University for generous 
hospitality during the GCD-V conference.

\section*{References}

\reference
Bell E.F.\ \& de Jong R.S., 2001, ApJ 550, 212

\reference 
Bottema R., 2002, A\&A 388, 809

\reference
Chabrier G., 2001, ApJ 554, 1274

\reference
Chabrier G., 2002, ApJ 567, 304

\reference 
Dale D.A., Giovanelli R., Haynes M.P., Campusano L.E.\ \&  Hardy E., 1999, 
AJ 118, 1489

\reference 
Efstathiou G., Lake G.\ \& Negroponte J., 1982, MNRAS 199, 1069

\reference 
Eke V.R., Navarro J.F.\ \& Steinmetz M., 2001, ApJ 554, 114

\reference 
Giovanelli R., Haynes M.P., Herter T.\ \& Vogt N.P., 1997, AJ 113, 53

\reference 
Girardi L., Bressan A., Bertelli G., Chiosi C., Groenewegen M.A.T.,
Marigo P., Salasnich B.\ \& Weiss A., 2002, A\&A 391, 195

\reference 
Kannappan S.J., Fabricant D.G.\ \& Franx M., 2002, AJ 123, 2358

\reference 
Kennicutt R.C., Tamblyn P.\ \& Congdon C.W., 1994, ApJ 435, 22 (KTC94)

\reference 
Kroupa P., 1998, in Brown Dwarfs and Extrasolar Planets, eds.\ R.\ Rebolo, 
E.L.\ Martin and M.R. Zapatero Osorio, ASP Conf.\ Series vol.~134, p.~483

\reference 
Kroupa P.\ \& Weidner C., 2003, ApJ 598, 1076

\reference 
Larson R.B., 1998, MNRAS 301, 569

\reference 
Moretti A., Portinari L.\ \& Chiosi C., 2003, A\&A 408, 431

\reference 
Navarro J.F.\ \& Steinmetz M., 2000a, ApJ 528, 607

\reference 
Navarro J.F.\ \& Steinmetz M., 2000b, ApJ 538, 477

\reference 
Pagel B.E.J., 1997, Nucleosynthesis and Chemical Evolution of Galaxies,
Cambridge University Press

\reference 
Portinari L.\ \& Chiosi C., 1999, A\&A 350, 827

\reference 
Portinari L., Chiosi C.\ \& Bressan A., 1998, A\&A 334, 505

\reference 
Portinari L., Sommer--Larsen J.\ \& Tantalo R., 2004a, 
{\mbox{MNRAS}} 347, 691 (PST)

\reference 
Portinari L., Moretti A., Chiosi C.\ \& Sommer--Larsen J., 2004b, 
ApJ in press (astro-ph/0312360)

\reference 
Rubin V.C., Burstein D., Ford W.K.\ \& Thonnard N., 1985, ApJ 289, 81 

\reference 
Salpeter E.E., 1955, ApJ 121, 161

\reference 
Scalo J.M., 1986, Fund.\ Cosmic Phys.\ 11, 1

\reference
Sommer--Larsen J.\ \& Dolgov A., 2001, ApJ 551, 608

\reference
Sommer--Larsen J., G\"otz M.\ \& Portinari L., 2003, ApJ 596, 47

\reference
Tinsley B.M., 1980, Fund. Cosmic Phys.\ 5, 287

\reference
Trott C.M.\ \& Webster R.L., 2002, MNRAS 334, 621

\reference
Vallejo O., Braine J.\ \& Baudry A., 2002, A\&A 387, 429

\end{document}